\documentclass[pra,twocolumn,showpacs,superscriptaddress,floatfix, reprint]{revtex4-1}

\usepackage[OT1]{fontenc}
\usepackage[usenames,dvipsnames]{color}
\usepackage[latin1]{inputenc}
\usepackage[english]{babel}
\usepackage{graphicx}
\usepackage{color}
\usepackage[Gray,squaren]{SIunits}
\usepackage{xspace}
\usepackage{ulem}
\usepackage{epstopdf}
\usepackage{amssymb,amsmath,verbatim,ulem}
\usepackage{dcolumn}
\usepackage{bm}
\usepackage{braket}
\usepackage{tabularx}
\usepackage{rotating}
\usepackage{hyperref}
\usepackage[caption=false]{subfig}

\newcommand{\RNum}[1]{\uppercase\expandafter{\romannumeral #1\relax}}

\begin{document}
\title{Microwave assisted optical waveguide in Rydberg atoms} 
\author{Nawaz Sarif Mallick}
\email{nawaz.phy@gmail.com}

\author{Tarak Nath Dey}
\email{tarak.dey@iitg.ac.in}
\affiliation{Department of Physics, Indian Institute of Technology
Guwahati, Guwahati, Assam 781039, India}

\date{\today}

\begin{abstract}
We theoretically demonstrate an efficient scheme to build a micro-wave (MW) assisted optical waveguide
in an inhomogeneously broadened
vapor medium that is made of active $^{87}$Rb atoms and inactive buffer gas atoms. We exploit the sensitive behaviour
of MW field coupled between highly excited Rydberg states to create distinctly responsive and tunable atomic waveguide.
The buffer gas induced collision further manipulates the features of the waveguide by widening the spatial transparency window
and enhancing the contrast of the refractive index.
We numerically solve Maxwell's equations to demonstrate diffractionless propagation
of 5 $\mu m$ narrow paraxial light beams of arbitrary mode to several Rayleigh lengths.
The presence of the buffer gas significantly enhances output intensity of the diffraction controlled light beam from 10\% to 54\%.
This efficient diffraction elimination technique has important applications in high-resolution imaging and
high-density optical communication.           
\end{abstract}


\maketitle

\section{Introduction} 
The ability to guide a narrow width optical beam holds promise for applications in
high-density optical communication \cite{Glezer:96} and high-resolution imaging \cite{Feng:06,Tsang17}. 
The main obstacle for realisation of narrow beam based optical technology
comes from diffraction and absorption of the medium  \cite{John:18}.
The divergence angle of a narrow beam is significantly larger as compared to
a broad beam due to its geometrical shape \cite{Eberly_2010}.
Consequently, the narrow beam encounters severe spatial distortions along
the transverse direction as it propagates a few  Rayleigh wavelengths distance through the medium.
Ultimately diffraction induced image blurring prevents its use in the important light based applications \cite{Glezer:96,Feng:06}.
Hence, the complete elimination of diffraction for the narrow width beam becomes a long-standing goal.

To achieve this goal, different methods based on Raman self-focusing technique \cite{Ladouceur:97,Stoyanov:14,Bulk1:OL},
and manipulation of refractive index \cite{GSA:PRA00,Xu:18,Akin:17,TND:PRA11} have been proposed
for bulk media \cite{Bulk1:OL,Xu:18} as well as for atomic vapor media \cite{GSA:PRA00,TND:PRA11}.
Suitable tailoring of refractive index along transverse direction leads to
the formation of waveguide like structure inside cold atomic media\cite{TND:OL09,Akin:17,Lida:PRA13}
and also in hot vapor cells \cite{Truscott:PRL99,GSA:PRA00,Lida:Sci17}.
Truscott $et$ $al.$ published their seminal paper establishing that an atomic vapor can produce a waveguide
which controls beam propagation without diffraction \cite{Truscott:PRL99}.
This experiment opened a floodgate for  numerous experiments
\cite{Truscott:PRL99,Friedmann:PRA05,Firstenberg:Nature09,Praveen:PRL09,Firstenberg:PRL09} in addition to
theoretical investigations \cite{GSA:PRA00,TND:OL09,TND:PRA11,Sandeep:PRA17,Truscott:PRA01,Lida:Sci17,Friedmann:PRA05,Lida:PRA13}. 
A suitably chosen spatial profile of the control field that creates spatial modulation
of refractive index enables weak optical beams to propagate through the medium without loss of generality.
Taking advantage of different spatial profiles of control field such as Gaussian \cite{TND:OL09,Friedmann:PRA05,DING20121954},
super-Gaussian \cite{Lida:Sci17,TND:OL09}
and different modes of Laguerre-Gaussian (LG$^{l}_{n}$) \cite{TND:PRA11,Sandeep:PRA17,RANJITACHANU2013150}
results in an undistorted probe beam dynamics. 
Off-resonance \cite{TND:PRA11,Lida:Sci17} or nearly resonance atomic transitions \cite{Sandeep:PRA17,Wang:08}
can also support   guiding and steering of optical  beams.
However, the optically written waveguide is based on normal atoms with low principal quantum number and
is often associated with considerable amount of absorption which limits lossless beam propagation to several Rayleigh lengths. 
Another drawback of normal atomic waveguide appears due to its lack of high contrast in refractive index between core and cladding
that fails to support narrow beam propagation.
These limitations can be overcome by exploiting the exaggerated optical properties of Rydberg atoms
with high principal quantum number  \cite{TF_1994}.
An atomic waveguide with narrow core and high contrast refractive index
is a very fundamental criteria for guiding a narrow beam with size of the order of $5  \mu m$.

In this article, we use highly excited Rydberg energy states of Rubidium atom to create high contrast and
narrow core optical waveguide. The inspiration of our work comes from recent experimental demonstration
of Shaffer {\it et al.} \cite{Shaffer:Nature12} in which MW field becomes highly responsive to the Rydberg
energy states \cite{Shaffer:Nature12,Vogt:OL18,KWAK2016168,Mallick_2017}.
Even a very weak MW field (8 $\mu V\: cm^{-1}$) is able to modify the probe response drastically 
\cite{Shaffer:Nature12}. We exploit this sensitive behavior of Rydberg energy states to create a highly efficient
and extremely tunable atomic waveguide. A high contrast refractive index modulation of probe can be obtained
by application of Bessel-Gaussian (BG) shaped MW beam which couples two highly excited Rydberg states
$\ket {30 D_{\frac{5}{2}}}$, $\ket {31 P_{\frac{3}{2}}}$ \cite{Vogt:OL18}. The desired spatial shape of the beam
either in MW or optical domain can be found experimentally \cite{Chu2015,Nowack2012,Fu:17,Li:17,GORI1987491}.
Note that the strength of the dipole-dipole interactions 
mediated through the residual occupation in Rydberg states are very small in the considered
model system \cite{Shaffer:Nature12,PhysRevA.94.043806,Vogt:OL18,PhysRevA.99.023832} and can be neglected safely.

Further, we assimilate inactive buffer gas atoms in addition to active Rydberg atoms inside
the vapor cell to enhance the efficiency of the waveguide \cite{Firstenberg:Nature09,Firstenberg:PRL09,Evers2014}.
The active atoms frequently collide with the buffer gas atoms in which the velocity of the former
alter from one velocity group to another velocity group. This velocity changing collision (VCC) leads
to the phenomena of Dicke narrowing \cite{Dicke:PRA07, Dicke:PRA08} in presence of buffer gas.
In this article, we exploit buffer gas induced VCC process in order to create a high contrast and efficient atomic waveguide.
The perspective of the current scheme is substantially unique from the preceding articles
in two ways \cite{Firstenberg:Nature09,Firstenberg:PRL09,Sandeep:PRA17}.
First, the key difference is the use of spatially modulated MW beam between the Rydberg states.
The spatial dependent MW BG beam generates a sharply varying fiber-like refractive index profile
which is tightly confined in the central region of the waveguide.
Second, the presence of buffer gas further manipulates the features of the waveguide
by widening the transparency window and enhancing the contrast of the refractive index profile. 
Therefore, the transmission of the weak diffraction controlled probe beam
at medium output enhances from 10\% to 54\% in the presence of buffer gas, unlike the results based on
the absorptive systems reported earlier \cite{Firstenberg:Nature09,Firstenberg:PRL09,Sandeep:PRA17}. 
Also the enhanced contrast in refractive index focuses the probe beam tightly towards the center of the waveguide.
Along with that the waveguide possesses an exclusive and handy feature.
Its absorption and refractive index profile are squeezed from both sides with changing order of the BG beam,
which makes this waveguide very efficient in guiding the probe beam of arbitrary width.
A narrow beam broadens much faster than the wide beam because the divergence angle is inversely proportional to the beam width.
Therefore, this high contrast and tunable atomic waveguide is essential for diffraction elimination from the narrow beam
of any arbitrary mode.

The paper is structured as follows. In section \ref{THEORETICAL},
we introduce the model configuration and describe the interaction of Rydberg atoms with the optical and MW fields
by a semi-classical density matrix formalism.
In section \ref{MW}, we point out the advantage of using a Rydberg atomic system over its normal counterpart
by studying the probe susceptibility. In section \ref{FAW}, we display
how the spatial structure of the MW beam permits us to build an optical waveguide.
In section \ref{Tun}, we discuss the tunability of the atomic waveguide.
Section \ref{Beam} demonstrates the propagation of weak probe field
having different beam profiles through the atomic waveguide.
Finally, we briefly conclude our work in section \ref{CONCLUSION}.

\section{THEORETICAL MODEL}
\label{THEORETICAL}
\begin{figure}
\begin{center}
\includegraphics[scale=0.3]{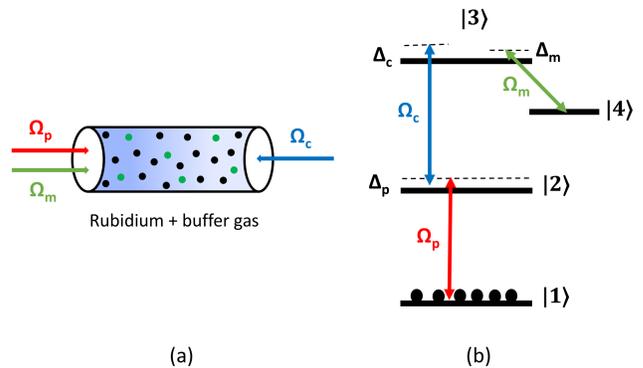}
\caption{(a) A simple illustration of the model system. The vapor cell contains active Rubidium atoms (black dots) and
inactive buffer gas atoms (green dots). Two counter-propagating optical fields $\Omega_p$, $\Omega_c$
and one MW field interact with the active atoms. 
(b) Schematic representation of the four level system. The energy levels have been realized in
$^{87}$Rb atomic vapor where $\ket1$=$\ket {5 S_{\frac{1}{2}}, F=2, m_{F}=2}$,
$\ket2$=$\ket {5 P_{\frac{3}{2}}, F=3, m_{F}=3}$, $\ket3$=$\ket {30 D_{\frac{5}{2}}, m_{J}=\frac{5}{2}}$,
$\ket4$=$\ket {31 P_{\frac{3}{2}}, m_{J}=\frac{3}{2}}$.}
\label{Figure1}
\end{center}
\end{figure}
\subsection{Model Configuration}
In this work, we study the collective behavior of active Rydberg atoms in the presence of inactive buffer gas atoms at room-temperature.
The geometry of the model system under consideration is shown in Fig. \ref{Figure1}(a) where two  counter propagating optical fields and MW field interact with the active $^{87}$Rb atoms. 
Figure \ref{Figure1}(b) shows four energy levels of active atoms which include one
metastable ground state $\ket1$ and three excited states $\ket2$, $\ket3$, $\ket4$ \cite{Shaffer:Nature12,Vogt:OL18}.   
The ground state $\ket1$=$\ket {5 S_{\frac{1}{2}}, F=2, m_{F}=2}$ is coupled to an excited state
$\ket2$=$\ket {5 P_{\frac{3}{2}}, F=3, m_{F}=3}$ by a weak probe field.
Two highly excited Rydberg states $\ket3=\ket {30 D_{\frac{5}{2}}, m_{J}=\frac{5}{2}}$ and
$\ket4$=$\ket {31 P_{\frac{3}{2}}, m_{J}=\frac{3}{2}}$ are coupled by a moderately intense
MW field of frequency 84.2 GHz \cite{Vogt:OL18}.
A strong control field connects two states $\ket2$ and $\ket3$.
The electric fields associated with the electro-magnetic (EM) radiations are described as
\begin{equation}\label{eq:1}
\vec{E}_{j}(\vec{r},t)=\hat e_{j}\mathcal E_{j}(\vec{r})e^{i(k_j z-\omega_j t)}+c.c.,	
\end{equation}
where $\mathcal E_{j}(\vec{r})$, $k_j$, $\omega_j$ and $\hat e_{j}$ are the slowly varying envelope, wave number, frequency
and unit polarization vector of the EM fields respectively.
The indices $j\in \{p,c, m\}$ refer to the probe, control and MW field.          
The EM fields only interact with the active atoms and the interaction can be expressed as a time-dependent Hamiltonian under the electric dipole approximation :
\begin{equation}\label{eq:2}
\begin{aligned}
H^{'}=&\hbar \omega_{21}\ket2\bra2+\hbar(\omega_{21}+\omega_{32})\ket3\bra3 \\
+&\hbar(\omega_{21}+\omega_{32}-\omega_{34})\ket4\bra4-\hbar\Omega_{p}e^{-i\omega_{p}t}\ket2\bra1 \\
-&\hbar\Omega_{c}e^{-i\omega_{c}t}\ket3\bra2-\hbar\Omega_{m}e^{-i\omega_{m}t}\ket3\bra4 + h.c.,\\
\end{aligned}
\end{equation}   
where $\Omega_p$, $\Omega_c$, $\Omega_{m}$ are the Rabi frequencies of the probe, control and MW fields respectively.
The expression of Rabi frequencies are  
\begin{equation}\label{eq:rabi}
\Omega_p=\frac{\vec{d}_{21}.\hat e_p}{\hbar} \mathcal E_{p},\quad \Omega_c=\frac{\vec{d}_{32}.\hat e_c}{\hbar} \mathcal E_{c}\quad \text{and}\quad \Omega_{m}=\frac{\vec{d}_{34}.\hat e_{m}}{\hbar} \mathcal E_{m}.
\end{equation}
In order to acquire the time-independent Hamiltonian, we execute the following unitary transformation
\begin{equation}
H=U^\dagger H^{'} U -i\hbar U^\dagger \frac{\partial U}{\partial t},
\end{equation}
where $U$ is defined as
\begin{equation}
U=e^{-i\left(\omega_{p}\ket2\bra2 + (\omega_{p}+\omega_{c})\ket3\bra3 + (\omega_{p}+\omega_{c}-\omega_{m})\ket4\bra4\right)t}.
\end{equation}
Now the Hamiltonian transforms into the following time-independent form
\begin{equation}\label{eq:2}
\begin{aligned}
H=&-\hbar\Delta_p \ket 2\bra 2-\hbar(\Delta_p+\Delta_c) \ket 3\bra 3 \\
&-\hbar(\Delta_p+\Delta_c-\Delta_{m}) \ket 4\bra 4 -\hbar \Omega_p \ket 2\bra 1 \\
&-\hbar \Omega_c \ket 3\bra 2 -\hbar \Omega_{m} \ket 3\bra 4 + h.c.
\end{aligned}
\end{equation}
where $\Delta_p=\omega_p-\omega_{21}$, $\Delta_c=\omega_c-\omega_{32}$, and $\Delta_{m}=\omega_{m}-\omega_{34}$, are the detuning
of the probe, control and MW fields respectively. The resonant transition frequency between energy levels
$\ket 1 \leftrightarrow \ket 2$, $\ket 2 \leftrightarrow \ket 3$ and $\ket 3 \leftrightarrow \ket 4$ are denoted by
$\omega_{21}$, $\omega_{32}$ and $\omega_{34}$, respectively.

\subsection{Dynamical Equations}
The dynamics of the active atoms inside the vapor cell are governed by the following Liouville's equation :
\begin{equation}\label{dynamics}
\dot \rho=-\frac{i}{\hbar}[H,\rho]+ \mathcal{L}_\rho 
\end{equation}
where the second term incorporates various radiative and non-radiative decay processes in the presence of buffer gas atoms.
The collisions between the active atoms and buffer gas atoms change the velocity distribution of the active atoms
inside the medium. These collisions also affect the phase coherence between the atomic energy levels
which modifies the life-time of the atomic coherence \cite{kernel86,Singh:OSA88,Novikova:05,Singh:PRA09,CHRAPKIEWICZ20141}.
The effect of such collisions can be included in the dynamical Eq. (\ref{dynamics}) by adding the following term
\cite{kernel86,Singh:OSA88,Singh:PRA09}
\begin{equation}\label{collision}
\begin{aligned}
&\left[\frac{\partial \rho_{jk} (v,t)}{\partial t}\right]_{collision}=-\gamma_{ph}(1-\delta_{jk})\rho_{jk}(v,t)\\
&-\Gamma_{jk}\rho_{jk}(v,t)+\int K(v^{\prime}\rightarrow v) \rho_{jk}(v^{\prime},t) dv^{\prime}
\end{aligned}
\end{equation}
In the above Eq. (\ref{collision}), $\Gamma_{jk}$ is known as velocity changing collision rate and
$\gamma_{ph}$ is the rate of collisional dephasing of the atomic coherence.           
The collision kernel, $K(v^{\prime}\rightarrow v)$ represents the probability density per unit time that
active atoms have when their velocity change from $v^{\prime}$ to $v$ as a result of collisions
with buffer gas atoms \cite{kernel86}.
For simplicity, the collision kernel can be written in terms of $\Gamma_{jk}$ as shown in the following expression
\begin{equation}
\begin{aligned}
&K(v^{\prime}\rightarrow v)=\Gamma_{jk} M(v),\\
&M(v)=\frac{1}{\sqrt\pi v_{th}}e^{-\frac{v^{2}}{v^{2}_{th}}}, \quad v_{th}=\sqrt{\frac{2k_B T}{m_{A}}},
\end{aligned}
\end{equation}
where $M(v)$ and $v_{th}$ are the Maxwell-Boltzmann velocity distribution along the $z$ direction and
most probable velocity of the active atoms of mass $m_{A}$ at a temperature $T$.
The spontaneous decay rates from the excited state $\ket j$, $(j \in 2,3,4)$ to
the ground state $\ket 1$, are denoted by $\gamma_{j1}$.
Note that the collision rates of different atomic states $\Gamma_{j1}$, $j \in 2,3,4$ are all similar in strength
and are indicated by $\Gamma_{21}\simeq \Gamma_{31}\simeq \Gamma_{41}$=$\Gamma_{c}$ \cite{Singh:PRA09}. 
In the considered model system, $k_{p}$ ($\simeq 2\pi \times 1.3 \:\mu m^{-1}$) is nearly equal to
$k_{c}$ ($\simeq 2\pi \times 2.0 \:\mu m^{-1}$) such that the
wave vector difference $|\delta \vec{k}|=|\vec{k}_{p}-\vec{k}_{c}|$ becomes minimal.
Further, we restrict our analysis for the moderate collision case in which $\Gamma_{c}$ is comparatively smaller than spontaneous decay rate ($\gamma_{21}$) and Doppler width ($\gamma_{d}$) $i.e.$ $\Gamma_{c} \ll \gamma_{21}, \gamma_{d}$
\cite{Singh:PRA09, Shaffer:Nature12}.
The degree of collisions can be realized experimentally by controlling the density of the buffer gas inside the vapor cell.
The collision induced $\Gamma_c$ significantly influences the absorptive and dispersive features of the atomic medium.

The following three coupled density matrix equations are sufficient for describing the dynamics of the active atoms in the buffer gas environment under weak probe approximation
\begin{equation}\label{eq:r}
\begin{aligned}
\dot{\rho}_{21}(v,t)=&-A_{21}(v)\rho_{21}(v,t)+i\Omega_p(\rho_{11}(v,t)-\rho_{22}(v,t)) \nonumber\\
&+i\Omega^*_{c}\rho_{31}(v,t), \nonumber\\
\dot{\rho}_{31}(v,t)=&-A_{31}(v)\rho_{31}(v,t)+i\Omega_c\rho_{21}(v,t)-i\Omega_p\rho_{32}(v,t) \nonumber\\
&+i\Omega_{m}\rho_{41}(v,t)+\Gamma_{31}M(v)\int \rho_{31}(v,t) dv, \nonumber\\
\dot{\rho}_{41}(v,t)=&-A_{41}(v)\rho_{41}(v,t)+i\Omega^{*}_{m}\rho_{31}(v,t)-i\Omega_p\rho_{42}(v,t) \nonumber\\
&+\Gamma_{41}M(v)\int \rho_{41}(v,t) dv, \nonumber
\end{aligned}
\end{equation}
where
\begin{equation}\label{eq:rr}
\begin{aligned}
A_{21}(v)=&i(\vec{k}_p.\vec{v}-\Delta_p)+\gamma_{21}+\Gamma_{21}+\gamma_{ph}, \nonumber\\
A_{31}(v)=&i\{(\vec{k}_p+\vec{k}_c).\vec{v}-(\Delta_p+\Delta_c)\}+\gamma_{31}+\Gamma_{31}+\gamma_{ph}, \nonumber\\
A_{41}(v)=&i\{(\vec{k}_p+\vec{k}_c-\vec{k}_{m}).\vec{v}-(\Delta_p+\Delta_c-\Delta_{m})\} \nonumber\\
& +\gamma_{41}+\Gamma_{41}+\gamma_{ph}. \nonumber
\end{aligned}
\end{equation}
The perturbative solution of the atomic coherence and population in the limits of weak probe approximation can be defined as
\begin{equation}
\begin{aligned}
\rho_{jk}&=\rho_{jk}^{(0)}+\Omega_p\rho_{jk}^{(1)}.\\
\end{aligned}
\end{equation}
The zeroth order solution in the absence of probe field is $\rho_{11}^{(0)}=M(v)$ \cite{Singh:PRA09}.  
The first order solutions in the presence of weak probe field can be obtained in the following forms
\begin{equation}\label{eq:r}
\begin{aligned}
\dot{\rho}_{21}^{(1)}(v,t)&=-A_{21}(v)\rho_{21}^{(1)}(v,t)+i\rho^{(0)}_{11}+i\Omega^*_{c}\rho_{31}^{(1)}(v,t), \nonumber\\
\dot{\rho}_{31}^{(1)}(v,t)&=-A_{31}(v)\rho_{31}^{(1)}(v,t)+i\Omega_c\rho_{21}^{(1)}(v,t) \nonumber\\
&+i\Omega_{m}\rho_{41}^{(1)}(v,t)+\Gamma_{31}M(v)\int \rho_{31}^{(1)}(v,t) dv, \nonumber\\
\dot{\rho}_{41}^{(1)}(v,t)&=-A_{41}(v)\rho_{41}^{(1)}(v,t)+i\Omega^{*}_{m}\rho_{31}^{(1)}(v,t) \nonumber\\
&+\Gamma_{41}M(v)\int \rho_{41}^{(1)}(v,t) dv. \nonumber
\end{aligned}
\end{equation}
The steady state response of the above coupled equations {\it i.e.,} $\dot{\rho}_{jk}^{(1)}=0$ can be used
for finding the analytical expression of 
first order atomic coherences $\langle\rho_{21}\rangle$, $\langle\rho_{31}\rangle$ and $\langle\rho_{41}\rangle$. 
We also incorporated the thermal agitation of the atom by performing the velocity averaging of the atomic coherence
\begin{equation}
\begin{aligned}
\langle\rho_{21}\rangle&=\int \rho^{(1)}_{21}(v) dv, \nonumber\\
&=i\Omega_P f_{10}+i\Omega^*_c\Gamma_{31}f_{3}\langle\rho_{31}^{(1)}\rangle \nonumber\\
&-\Omega^*_c\Omega_{m}\Gamma_{41}f_{1}\langle\rho_{41}^{(1)}\rangle, \\
\langle\rho_{31}^{(1)}\rangle&=\int \rho^{(1)}_{31}(v) dv, \nonumber\\
&=\Omega_p\Omega_c \Bigg[\frac{|\Omega_{m}|^{2}f_{1}L_{2}-f_{3}L_{3}}{L_{1}L_{3}+|\Omega_{m}|^{2}L_{2}L_{4}}\Bigg], \nonumber\\
\langle\rho_{41}^{(1)}\rangle&=\int \rho^{(1)}_{41}(v) dv, \nonumber\\
&=-i\Omega_p\Omega_c\Omega^{*}_{m} \Bigg[\frac{f_{1}L_{1}+f_{3}L_{4}}{L_{1}L_{3}+|\Omega_{m}|^{2}L_{2}L_{4}}\Bigg],
\end{aligned}
\end{equation}
where
\begin{equation}
\begin{aligned}
L_{1}=&1+|\Omega_c|^2 \Gamma_{31}f_{8}-\Gamma_{31}f_{5}, \nonumber \\
L_{2}=&\Gamma_{41}f_{4}-|\Omega_c|^2 \Gamma_{41}f_{6}, \nonumber \\
L_{3}=&1-|\Omega_c|^2 |\Omega_{m}|^2 \Gamma_{41}f_{9} \nonumber \\
&-\Gamma_{41}f_{2}+|\Omega_{m}|^2 \Gamma_{41}f_{7}, \nonumber \\
L_{4}=&\Gamma_{31}f_{4}-|\Omega_c|^2 \Gamma_{31}f_{6}. \nonumber
\end{aligned}
\end{equation}
The expression of the spatial functions $f_{j}$, ( $j \in$ 1,2,3,$\dots$ 10) are shown in the appendix section (\ref{appendix}). 
Finally, the velocity averaged linear probe susceptibility, $\langle\chi_{21}\rangle$ at frequency $\omega_p$ can be written as
\begin{equation}
\begin{aligned}
\langle\chi_{21}\rangle=\frac{\mathcal N |\vec{d}_{21}|^2}{\hbar \Omega_p} \langle\rho_{21}\rangle
\end{aligned}
\end{equation}
where $\mathcal N$ is the atomic density.
\begin{figure}[t]
\begin{center}
\includegraphics[scale=0.4]{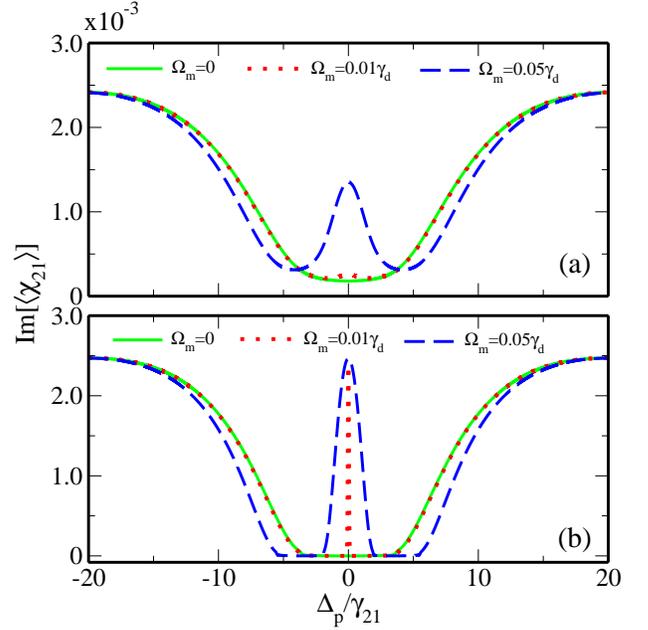}
\caption{Probe absorption profile, Im[$\langle\chi_{21}\rangle$] as a function of probe laser detuning ($\Delta_p$)
for (a) normal (b) Rydberg atomic system. EIA lineshapes for three different values of MW field ($\Omega_{m}$)
demonstrate the MW field sensitivity. Decay rate for (a) $\gamma_{31}\simeq\gamma_{41}\simeq\gamma_{21}$ and
for (b) $\gamma_{31}=2\pi\times 1.0\times 10^{3}$ Hz, $\gamma_{41}=2\pi\times 0.5\times 10^{3}$ Hz. 
The other parameters are $\Omega_c=0.3\gamma_d$, $\Delta_c=0$, $\Delta_{m}=0$, $\Gamma_{c}=0$, T=300K,
$\mathcal N=5\times 10^{10}$ atoms/cm$^3$, $\gamma_{d}=2.53\times 10^{9}$ Hz, $\gamma_{21}=2\pi\times 6.1\times 10^{6}$ Hz,
$\gamma_{ph}\approx 1\times 10^{3}$ Hz.}
\label{Figure2}
\end{center}
\end{figure}
\section{Microwave field sensitivity} 
\label{MW}
In this section, we distinguish the advantage of using Rydberg atomic system over its normal counterpart
by studying the probe susceptibility in the absence and presence of the MW field.
The normal counterpart here refers to a low quantum number atomic system with
$\ket3$=$\ket {5 D_{5/2}, m_{J}=5/2}$, $\ket4$=$\ket {6 P_{3/2}, m_{J}=3/2}$. 
In Rydberg atomic system, MW field couples two Rydberg states with high principal quantum number as shown in Fig. \ref{Figure1}(b).
Whereas MW field connects two states with very low principal quantum number for normal atomic system.
In Fig. \ref{Figure2}(a) and \ref{Figure2}(b), we compare the probe absorption lineshape in case of
normal as well as Rydberg atomic system for three different values of MW field
intensity $\Omega_{m}$=0, 0.01$\gamma_{d}$, 0.05$\gamma_{d}$.
In absence of MW field ($\Omega_{m}$=0), both the normal and Rydberg atomic systems display
electromagnetically induced transparency (EIT) \cite{Jenkins:19,Sharma:19}
under two-photon resonance condition ${\it i.e.,}$ $\Delta_{p}=\Delta_{c}$ as shown by solid green curves in Fig. \ref{Figure2}.
We have observed that the Rydberg system offers complete flat transparency window (no absorption)
due to the low decay rate ($\sim$ KHz) of the Rydberg states as compared to a shallow window
in the normal atomic states (decay rate $\sim$ MHz) \cite{agarwal_2012}.    
We next study how a weak MW field ($\Omega_{m}$=0.01$\gamma_{d}$) drastically modifies the probe response
in Rydberg systems which is distinct from a normal system.
It is clear from \ref{Figure2}(b) that a complete sharp electromagnetically induced absorption (EIA)
peak \cite{Hamedi:17,Bankim:17} is present in the Rydberg system.
On the other hand, the EIA peak in a normal atomic system just builds up as shown by the dotted red curve
in Fig. \ref{Figure2}(a).
With the increase of MW field power ($\Omega_{m}$=0.05$\gamma_{d}$), Rydberg EIA peak experiences power broadening,
while the normal EIA peak is still growing towards its maximum peak value.
These observations clearly confirm that Rydberg energy states are strongly responsive to the MW field
unlike the normal atomic states \cite{Shaffer:Nature12}.
We exploit this responsive behavior of MW field in Rydberg atomic systems to create
a highly tuneable atomic waveguide. 
The effective modulation of spatial susceptibility due to spatial structure of the MW beam holds the main essence
behind the formation of waveguide inside the Rydberg atomic system.

\begin{figure}[t]
\begin{center}
\includegraphics[scale=0.23]{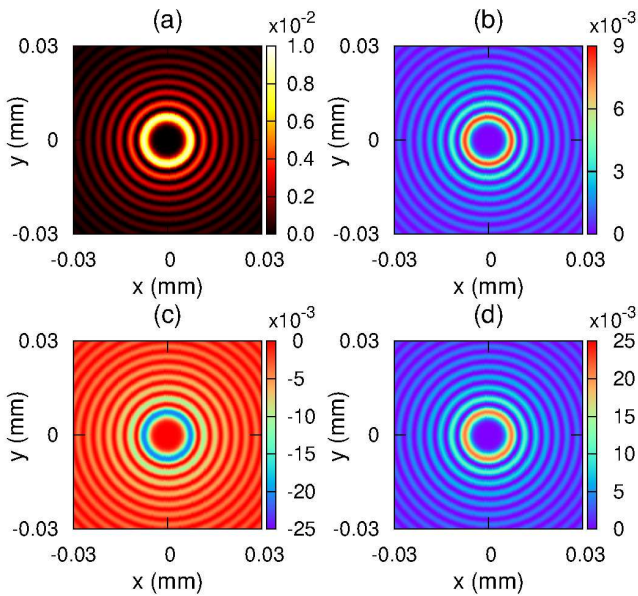}
\caption{(a) Input ring-shaped Bessel-Gaussian (BG) MW beam, (b) Fiber-like absorption profile tightly confined
in the central region of transverse position (x,y) with $\Delta_p=\pm 0.001\gamma_{21}$,
(c) Blue shifted detuning ($\Delta_p=0.001\gamma_{21}$)  and (d) Red shifted detuning ($\Delta_p=-0.001\gamma_{21}$)
depict waveguide like and anti-waveguide like refractive index profile in the transverse position $(x,y)$ respectively.
The parameters are $\Omega_c=0.3\gamma_d$, $\Omega_{m}=0.05\gamma_d$, $\omega_c=60\mu m$, $\omega_{m}=60\mu m$, $n=7$.
All other parameters are same as in \ref{Figure2}.}
\label{waveguide}
\end{center}
\end{figure}
\section{Formation of atomic waveguide}
\label{FAW}
We now investigate how the spatial structure of the MW beam permits us to build an optical waveguide inside the atomic medium.
A waveguide like refractive index can be formed by considering the transverse profile of the MW beam to be
Bessel-Gaussian (BG) together with a Gaussian (HG$^{0}_{0}$) shaped control beam.
The spatial shape of the MW BG beam in cylindrical coordinate can be expressed as \cite{Eberly_2010},
\begin{gather}
\begin{aligned}
\Omega_{m}(r,\phi,z=0) &=\Omega^{0}_{m} \mathcal J_{n}(ar) e^{in\phi} e^{-\frac{r^{2}}{w^{2}_{m}}} \: ;\\
&r=\sqrt{x^2+y^2},\quad \phi=\tan^{-1}\left(\frac{y}{x}\right).
\end{aligned}
\label{LG} 
\raisetag{15pt}
\end{gather}
The input amplitude of BG beam is denoted by $\Omega^{0}_{m}$. The $n$th-order Bessel function of first kind
is defined as $\mathcal J_{n}(ar)$, where $a$ is a scale factor. The minimum beam waist of Gaussian part is $w_{m}$.
The input spatial inhomogeneous intensity distribution of BG beam of order $7$ is shown in Fig. \ref{waveguide}(a).
It is clear from Fig. \ref{waveguide}(a) that the MW BG beam consists of many concentric rings which
has zero intensity at the central region whereas maximum intensity occurs in the ring-shaped regions.
Therefore, the MW BG beam together with Gaussian control beam can be used to obtain the desired spatial refractive
profile of the probe beam as evident from the following discussion.
Fig. \ref{waveguide}(b) shows the transverse variation of the probe absorption.
A complete transparency window exists at the core because of dominant characteristics of control beam over the MW field.
The diminishing intensity of the control beam towards the wing region yields absorption at the cladding.
Simultaneously, the MW field gains maximum intensity in the bright ring which causes high probe absorption due to EIA
as shown in Fig. \ref{waveguide}(b). Hence, considering a suitable spatial structure of the two fields, allows us
to achieve probe transparency at the core and opaqueness at cladding, at both resonance condition as well as near resonance
situation. 
Fig. \ref{waveguide}(c) displays that the refractive index attains a maximum value at EIT dominant region and
forms the core of the atomic waveguide. 
The cladding section of the waveguide can be cast by EIA, since EIT is ineffective in the ring-shaped region.  
The induced waveguide structure consists of refractive variation between core and cladding accompanied with a small width of
the core.
Hence the spatial response of the medium for the probe field exhibits waveguide like structure
at blue detuned condition whereas at red detuned condition, it changes to the anti-waveguide like structure
as shown in Fig. \ref{waveguide}(d). 
In both the cases, the core region of the atomic wave guide display minimum absorption  as shown in Fig. \ref{waveguide}(b). 
Finally, we have chosen the blue detuned probe field condition, {\it i.e.,} $\Delta_{p}=0.001\gamma_{21}$
for efficient guiding of various narrow Gaussian and Hermit Gaussian beams with arbitrary modes.

\begin{figure}[t!]
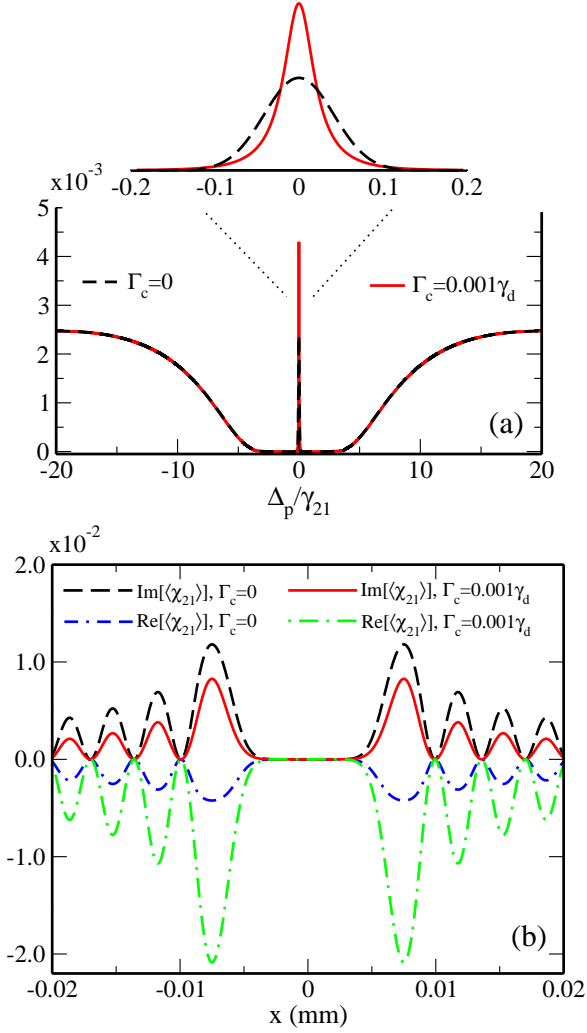

\begin{center}
\includegraphics[scale=0.46]{Figure4a.eps}
\includegraphics[scale=0.45]{Figure4b.eps}
\caption{(a) Probe absorption is plotted as a function of probe field detuning in the presence and absence of VCC.
Inset zoom figure shows the Dicke narrowing and enhancing of the EIA peak due to $\Gamma_{c}$.
(b) Real and imaginary part of the susceptibility is plotted as a function of transverse position $x$ at $y=0$ plane for two different
values of $\Gamma_{c}$. 
All other parameters are same as in \ref{Figure2}.}
\label{Figure4}
\end{center}
\end{figure}
\begin{figure}[b]
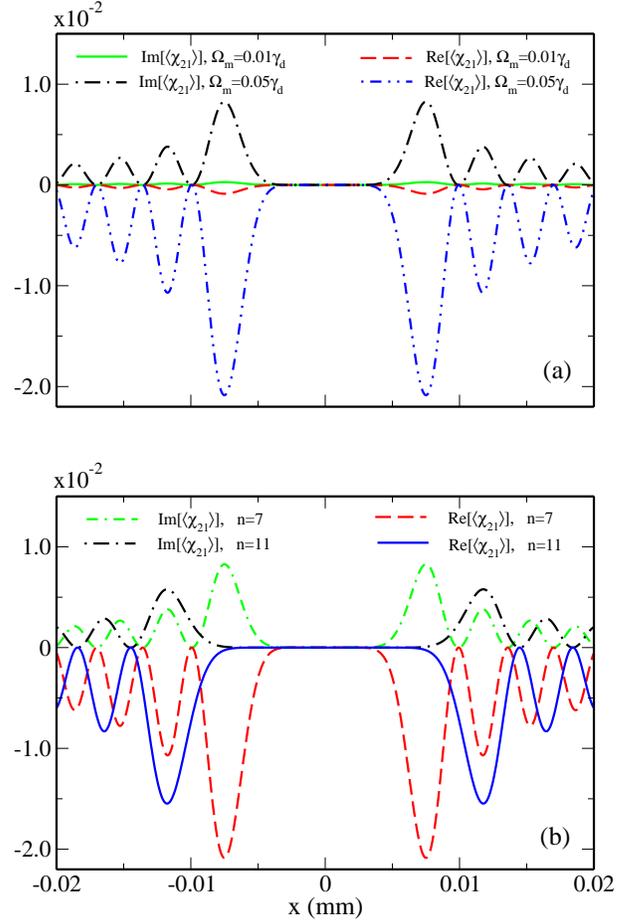

\begin{center}
\includegraphics[scale=0.37]{Figure5a.eps}
\includegraphics[scale=0.37]{Figure5b.eps}
\caption{Medium susceptibility along the transverse position $x$ at $y=0$ plane is plotted for two different values of
(a) MW field intensity and (b) BG beam order.  
The parameters are $\Omega_{c}=0.3\gamma_d$, $\Delta_p=0.001\gamma_{21}$, $\Gamma_{c}=0.001\gamma_d$.
All other parameters are same as in \ref{Figure2}.}
\label{Figure5}
\end{center}
\end{figure}


\section{Tunability of the waveguide}
\label{Tun}
Next, we discuss how the buffer gas induced collision significantly manoeuvres the features of the atomic waveguide
along the transverse direction in the presence of MW field ($\Omega^{0}_{m}$).
In order to comprehend the reasons behind these manipulations of atomic waveguide, we plot
the probe absorption lineshape as a function of $\Delta_p$ in the absence and presence of VCC as shown in Fig. \ref{Figure4}(a).
The MW induced EIA peak in Fig. \ref{Figure4}(a) can be enhanced by a notable amount in the presence of VCC.
Along with that the EIA lineshape manifests Dicke like narrowing due to the collision process as displayed
in the inset of Fig. \ref{Figure4}(a). Fig. \ref{Figure4}(b) illustrates that the contrast of refractive index
significantly enhances in the presence of VCC. The slope of VCC induced refractive index profile is remarkably
sharp which makes this waveguide more efficient in guiding the narrow probe beam in comparison to that of a Kerr
field induced waveguide \cite{Sandeep:PRA17}.
Further, the transparency window of the waveguide becomes much wider in the presence of the buffer gas.
As a result, a narrow probe beam propagates through the waveguide without significant loss of intensity.
These VCC induced Dicke narrowing and enhancing of the EIA peak distinctly facilitates the waveguide characteristics.

Now, we delineate how the MW field intensity and BG beam order effect the induced atomic waveguide.
In Fig. \ref{Figure5}(a), we plot imaginary and real part of the probe susceptibility along the transverse position for two
different values of MW field intensity. We observe that the difference of refractive index between the core and cladding
of the waveguide is enhanced significantly with the increase of MW intensity.
Fig. \ref{Figure5}(b), depicts the variation of absorption and refractive index profile with
the order of BG beam. A lower order BG beam squeezes the absorption profile and refractive index of the waveguide
from both sides.  
These unique features create sharply varying and confined refractive index profiles which makes the waveguide
very efficient in guiding the light beam.
It is familiar that a narrow beam broadens more rapidly than a wide beam due to optical diffraction because divergence angle
($\Theta=2\lambda_{p}/\pi w_{p}$) is inversely proportional to the beam width $w_{p}$ \cite{Eberly_2010}.
Hence, this high-contrast and squeezed waveguide is highly desired in order to remove diffraction from the narrow beam.
 
\section{Beam propagation through the waveguide}
\label{Beam}
Next, we study the propagation of weak probe field having different beam profiles through the atomic waveguide.
The beam propagation dynamics is governed by the Maxwell's wave equations \cite{TND:PRA11,Sandeep:PRA17}.
Under slowly varying envelope and paraxial wave approximation, Maxwell's wave equations for the
probe beam transform into the following form
\begin{equation}\label{eq:propagation}
\begin{aligned}
\frac{\partial \Omega_p}{\partial z}=\frac{i}{2{k_p}}\nabla_{\bot}^{2} \Omega_p + 2 i \pi k_p \langle \chi_{21} \rangle \Omega_p.
\end{aligned}
\end{equation}
\begin{figure}
\begin{center}
\includegraphics[scale=0.4]{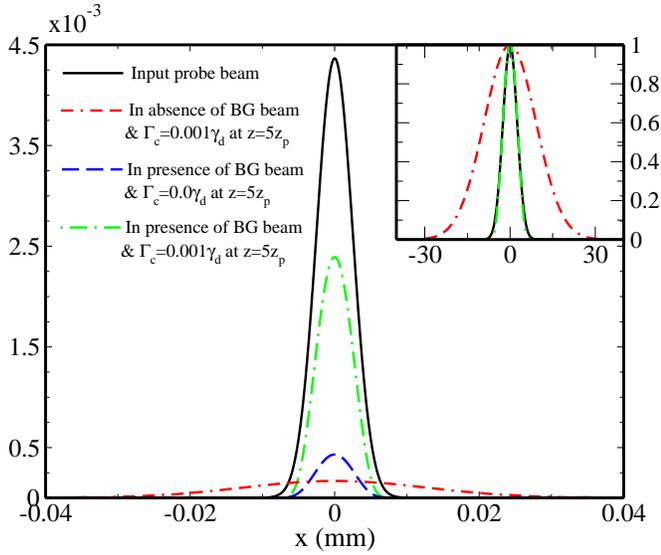}
\caption{Narrow Gaussian beam ($\omega_p=5\mu m$) propagation through the atomic medium
in presence and absence of MW BG beam and buffer gas. Inset figure shows the normalised intensity profile
of the probe beam in presence and absence of MW BG beam.
All other parameters are same as in figure \ref{Figure5}.}
\label{GausPropa}
\end{center}
\end{figure}
In Eq. (\ref{eq:propagation}), second order partial derivative in the $xy$ plane $i.e.$
$\nabla_{\bot}^{2}=\left( \partial^2 /\partial x^2 + \partial^2 /\partial y^2 \right)$ incorporates
inherent optical diffraction of the probe beam.
The last term of Eq. (\ref{eq:propagation}) is the contribution of linear and nonlinear optical effects which includes
MW and buffer gas induced absorption and refractive index profile of the medium in order to suppress the diffraction.
Note that we have considered MW Bessel-Gaussian (BG) beam which can propagate through
the atomic medium without diffraction. Also, the Rayleigh length of control beam is much larger than
the length of the atomic medium. Therefore, we can safely neglect the diffraction of the control beam.
We adopt split-step Fourier method (SSFM) to obtain the numerical solution of Eq. (\ref{eq:propagation})
and demonstrate the effect of spatially varying absorption and refractive index on probe beam dynamics.
First, we study the propagation dynamics of a Gaussian (HG$^{0}_{0}$) probe beam.
The width of the probe beam is 5 $\mu m$ which remains within the limit of
paraxial wave approximation, $\lambda_{p}/2\pi w_{p} < 0.1$ \cite{Agrawal:79,Sandeep:PRA17}.
The propagation dynamics of the narrow Gaussian shaped probe beam through the high contrast atomic waveguide
is shown in Fig. \ref{GausPropa}. The input and output intensity profile of the probe beam
in the presence and absence of the MW beam and buffer gas are illustrated clearly.
In absence of MW BG beam, the probe beam suffers diffraction induced broadening along with large absorption as
shown with double-dashed-dot red curve in Fig. \ref{GausPropa}.
The diffraction of the probe beam is drastically reduced in the presence of
MW BG beam. We notice that the output intensity of the probe beam decreases below 10\% in absence of
buffer gas after the propagation of $z=5 z_p$.
The situation changes in the presence of buffer gas, and the transmissivity of the diffraction
controlled probe beam enhances over 54\% for the same propagation distance ($z=5 z_p$).
This efficient beam propagation without any diffraction is possible due to the presence of
high-contrast tunable optical waveguide in buffer gas medium.

\begin{figure}
\begin{center}
\includegraphics[scale=0.95]{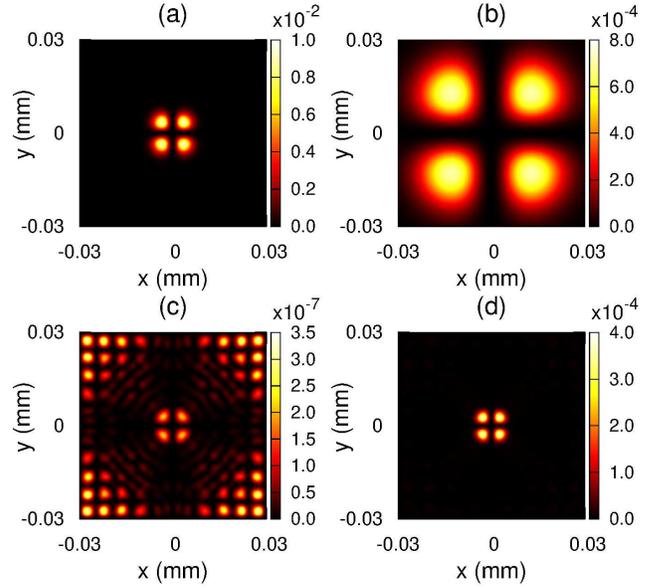}
\caption{(a) Intensity profile of the probe beam (HG$^{1}_{1}$) at z=0.
(b) Intensity profile of the probe beam (HG$^{1}_{1}$) in absence of MW BG beam at $z=5$$z_p$.
Intensity profile of the probe beam (HG$^{1}_{1}$) in presence of MW BG beam after propagation
over 5 Rayleigh lengths (z=5$z_p$) through the atomic medium (c) without and (d) with the buffer gas environment.
All other parameters are same as in figure \ref{GausPropa}.}
\label{HG11}
\end{center}
\end{figure}
In order to prove the robustness of the atomic waveguide, we also demonstrate diffraction-less propagation of
arbitrary Hermite-Gaussian (HG$^{v}_{u}$) modes. A further reason to choose HG$^{v}_{u}$ modes of narrow width
is its direct application in super-resolution imaging \cite{Zhou:18,Tsang17}. 
The spatial profile of different  HG$^{v}_{u}$ modes at its point of entering the medium $(z=0)$ is given by
\begin{equation}\label{Hermite-Gaussian}
\Omega_{p}(x,y,0)=\Omega^{0}_{p} H_{u}\left(\frac{\sqrt{2}x}{w_{p}} \right) H_{v}\left(\frac{\sqrt{2}y}{w_{p}} \right) e^{-\frac{x^2+y^2}{w^{2}_{p}}},
\end{equation}
where $H_u$ and $H_v$ are the Hermite polynomials of order $u$ and $v$ respectively.
For demonstration, we propel the Hermite-Gaussian probe beam of mode $u=1$, $v=1$ through the waveguide.
The intensity profile of the HG$^{1}_{1}$ mode at $z=0$ is shown in Fig. \ref{HG11}(a). 
The diffraction of the beam in absence of MW BG beam and buffer gas is displayed in Fig. \ref{HG11}(b).
The presence of BG beam and buffer gas eliminates the diffraction completely as clearly shown in Fig. \ref{HG11}(d).
However, Fig. \ref{HG11}(c) shows that in the absence of buffer gas medium, the waveguide fails to guide
HG$^{1}_{1}$ beam due to the lack of sharply varying high contrast refractive index profile.
The transverse structure of the MW beam and buffer gas play the important role in guiding the weak probe beam of
narrow width and arbitrary modes.  

\section{Conclusion}
\label{CONCLUSION}
In conclusion, we have demonstrated an efficient scheme to generate MW assisted optical waveguide in an inhomogeneously broadened
vapor medium that is made of active $^{87}$Rb atoms and inactive buffer gas atoms.
The sensitive behaviour of a MW field coupled between two highly excited Rydberg states of Rb atoms allow us to create a
responsive atomic susceptibility. The structured MW BG beam and Gaussian control beam together build
an optical waveguide with amenable fiber like refractive index profile.
The presence of buffer gas induced collision further manipulates the features
of the waveguide by widening the spatial transparency window and enhancing the contrast of the refractive index.
Changing the order of the MW BG beam squeezes the high contrast waveguide from both sides, which can successfully
guide the probe beam of narrow width.
We numerically solve Maxwell's equations to demonstrate diffractionless propagation
of narrow paraxial light beams of arbitrary modes such as Gaussian, Hermite-Gaussian HG$^{1}_{1}$ to several Rayleigh lengths.
The output intensity of the diffractionless light significantly enhances in the presence of a buffer gas.
This efficient technique to eliminate diffraction from narrow light beams
have important applications in high-density optical communication \cite{Glezer:96} and high-resolution imaging
\cite{Feng:06,Zhou:18,Tsang17}.\\

\section*{APPENDIX}
\begin{equation}\label{appendix}
\begin{aligned}
f_{1}&=\int \frac{M(v)}{f_{D}(v)}dv, \quad f_{2}=\int \frac{M(v)}{A_{41}(v)}dv, \nonumber\\
f_{3}&=\int \frac{A_{41}(v)M(v)}{f_{D}(v)}dv, \nonumber\\
f_{4}&=\int \frac{M(v)}{A_{31}(v)A_{41}(v)+|\Omega_{m}|^{2}}dv, \nonumber\\
f_{5}&=\int \frac{A_{41}(v)M(v)}{A_{31}(v)A_{41}(v)+|\Omega_{m}|^{2}}dv, \nonumber\\
f_{6}&=\int \frac{A_{41}(v)M(v)}{\left(A_{31}(v)A_{41}(v)+|\Omega_{m}|^{2}\right)f_{D}(v)}dv, \nonumber\\
f_{7}&=\int \frac{M(v)}{\left(A_{31}(v)A_{41}(v)+|\Omega_{m}|^{2}\right)A_{41}(v)}dv, \nonumber\\
f_{8}&=\int \frac{A_{41}^{2}(v)M(v)}{\left(A_{31}(v)A_{41}(v)+|\Omega_{m}|^{2}\right)f_{D}(v)}dv, \nonumber\\
f_{9}&=\int \frac{M(v)}{\left(A_{31}(v)A_{41}(v)+|\Omega_{m}|^{2}\right)f_{D}(v)}dv, \nonumber\\ \quad \quad \quad
f_{10}&=\int \frac{M(v)\left(A_{31}(v)A_{41}(v)+|\Omega_{m}|^{2}\right)}{f_{D}(v)}dv, \nonumber\\
\end{aligned}
\end{equation}
where
\begin{equation}
f_{D}(v)=A_{21}(v)\left(A_{31}(v)A_{41}(v)+|\Omega_{m}|^{2}\right)+A_{41}(v)|\Omega_c|^{2}. \nonumber
\end{equation}

\section*{Acknowledgment}
T.N.D. and N.S.M. gratefully acknowledge funding by the Science and Engineering Research Board (Grant No. CRG/2018/000054).
N.S.M thanks Prof Wenhui Li for useful discussions.

\bibliography{reference}
\end{document}